\documentclass[prd,10pt,aps,twocolumn,nofootinbib,showpacs,floatfix,amssymb,amsmath]{revtex4-1}
\usepackage{slashed,rotating}

\newcommand{\beqn}{\begin{eqnarray}}
\newcommand{\eeqn}{\end{eqnarray}}
\newcommand{\eq}[1]{(\ref{#1})}
\newcommand{\bl}{{\biggl|}}

\newcommand{\cL}{{\cal L}}

\newcommand{\Z}{{\mathbb{Z}}}

\begin{document}

\title{Permanently rotating devices: extracting rotation from quantum vacuum fluctuations?}
\author{M. N. Chernodub}\email{maxim.chernodub(at)lmpt.univ-tours.fr; \\ On leave from ITEP, Moscow, Russia.}
\affiliation{CNRS, Laboratoire de Math\'ematiques et Physique Th\'eorique, Universit\'e Fran\c{c}ois-Rabelais Tours,\\ F\'ed\'eration Denis Poisson, Parc de Grandmont, 37200 Tours, France}
\affiliation{Department of Physics and Astronomy, University of Gent, Krijgslaan 281, S9, B-9000 Gent, Belgium}

\begin{abstract}
We propose a set of devices of simple geometrical design which may exhibit a permanent rotation due to quantum (vacuum) fluctuations. These objects -- which have no moving parts -- impose certain boundary conditions on quantum fluctuations thus affecting their vacuum energy similarly to the standard Casimir effect. The boundary conditions are chosen in such a way that the vacuum energy for a static device is larger compared to the energy of the vacuum fluctuations in a state when the device rotates about a certain axis. The optimal frequency of rotation is determined by geometry and moment of inertia of the device. We illustrate our ideas in a vacuum of a massless scalar field theory using simplest Dirichlet-type boundary conditions. We also propose an experimental setup to verify the existence of the rotational vacuum effect.
\end{abstract}


\date{March 28, 2011}

\maketitle

Classically, the nonrelativistic kinetic energy of rotation of a rigid body around a principal axis of inertia is a quadratic function on the angular frequency\footnote{Negative and positive frequencies correspond to clockwise and counterclockwise directions of rotation, respectively.} $\Omega \equiv \pm |{\boldsymbol \Omega}|$:
\beqn
E^{\mathrm{cl}}_{\mathrm{rot}}(\Omega) = \frac{I \Omega^{2}}{2}\,,
\label{eq:E:class}
\eeqn
where  $I$ is the corresponding principal moment of inertia. The  most energetically favorable state corresponds to the static case, $\Omega = 0$.

In a quantum world, elementary particles have an important intrinsic degree of freedom, called spin $\boldsymbol s$, which corresponds to an intrinsic angular momentum of particles. A physical interpretation of the spin corresponds to a permanent rotation of the elementary particles around their own axis. Spin is quantized in terms of the Planck constant\footnote{We also use the convention $\hbar = c = 1$ unless stated otherwise.}, $|{\boldsymbol s}| = \sqrt{s(s+1)} \hbar$, where $s=0$ (scalar particles), $s=1/2$ (fermions), $s=1$ (vector bosons) etc.  

Can we construct a {\emph{macroscopic}} object which would be rotating permanently? Or, in other words, can a lowest energy state of a microscopically large object -- made of the Avogadro--scale  constituent particles ($N \sim 10^{23}$) -- correspond to a permanent rotation of this object around a certain axis? In order for this to happen, the rotational energy $E \equiv E_{\mathrm{rot}}(\Omega)$ of the object should have a minimum at a nonzero angular frequency $\Omega = \Omega_{\mathrm{opt}} \neq 0$ contrary to its classical analogue~\eq{eq:E:class}. 

\begin{figure}[!thb]
\begin{center}
\begin{tabular}{cc}
\includegraphics[scale=0.3,clip=false]{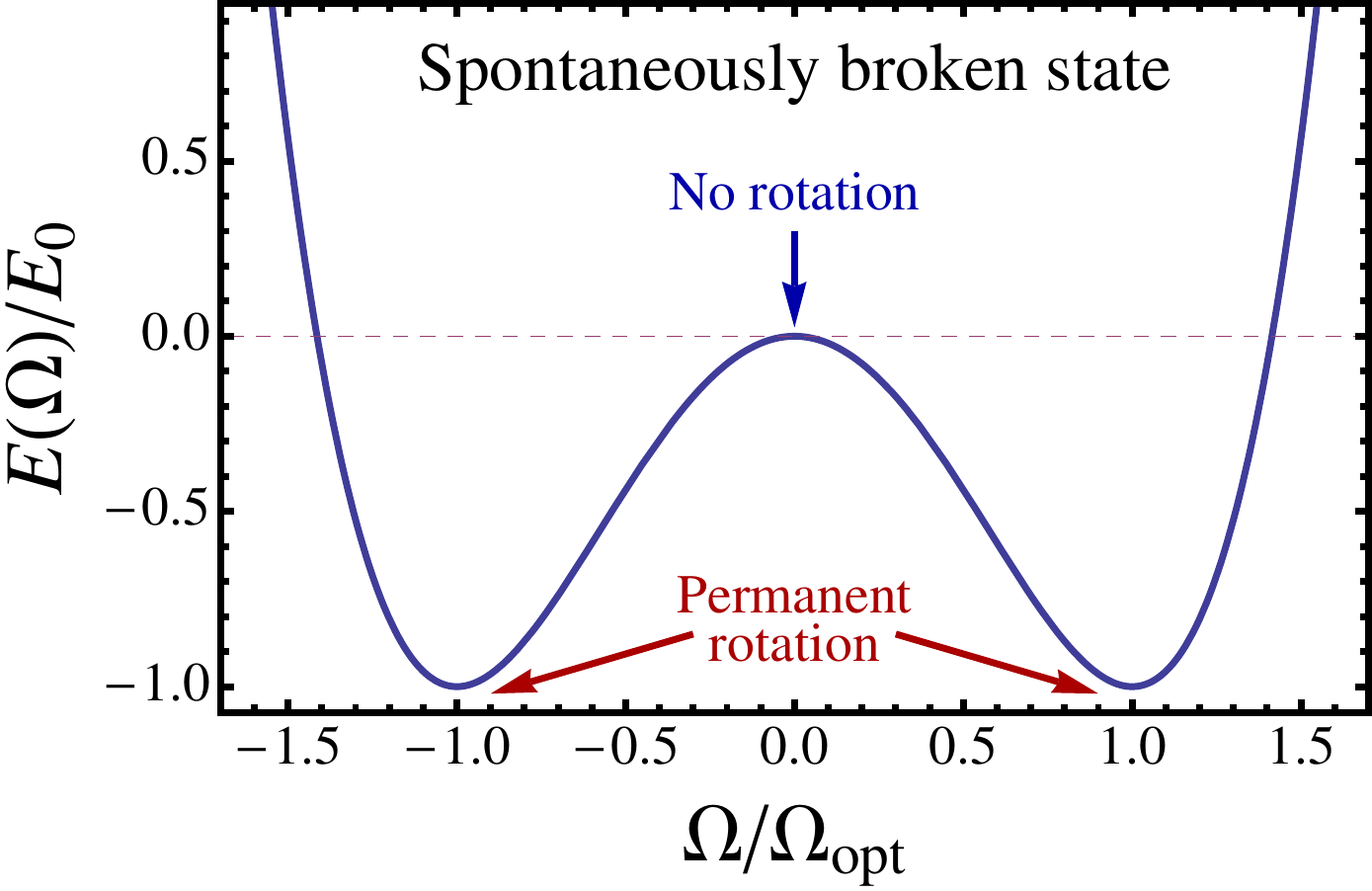} & \\[-28.5mm]
& \includegraphics[scale=0.27,clip=false]{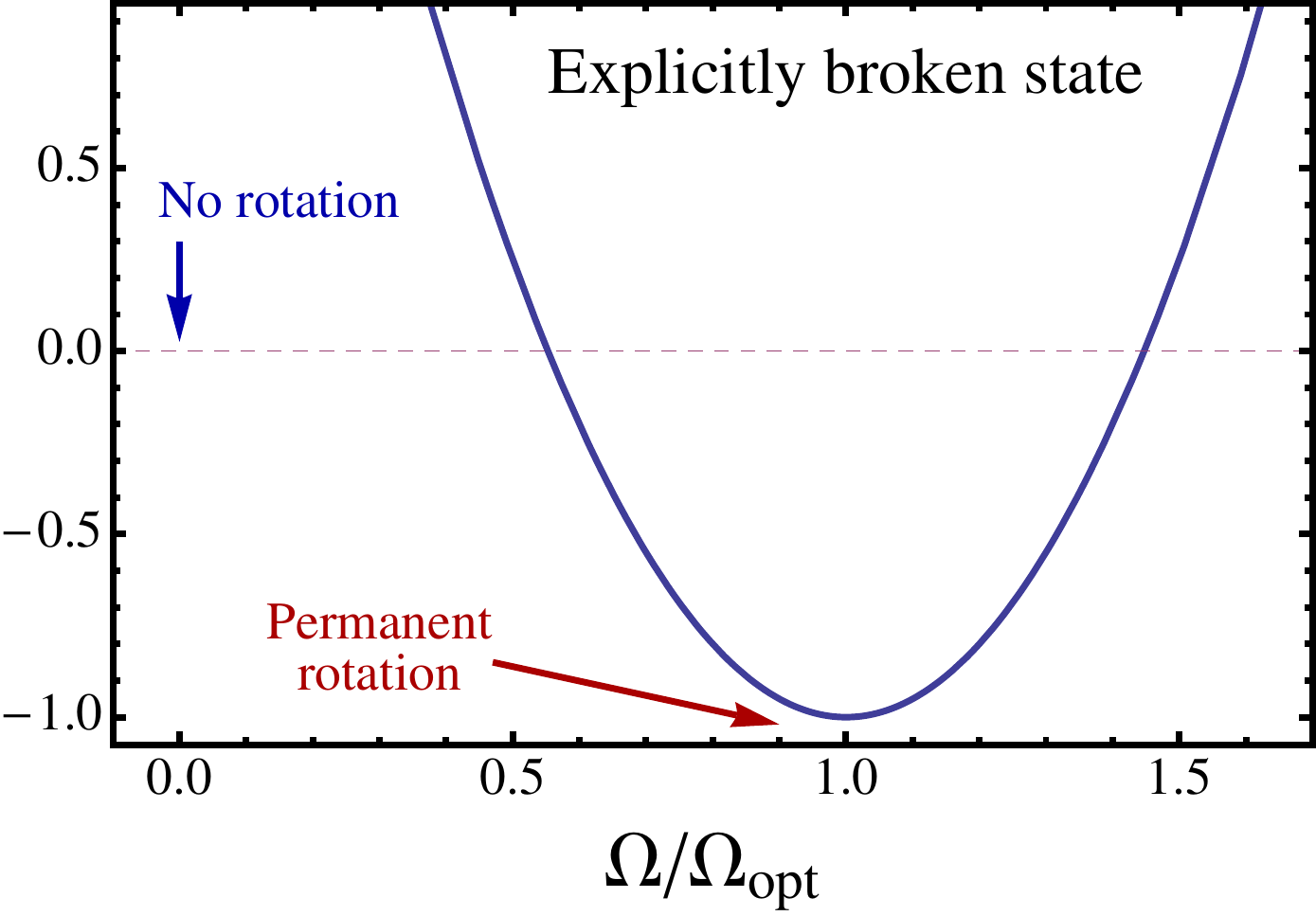} \\
\qquad (a) & \quad (b)
\end{tabular}
\end{center}
\vskip -5mm
\caption{Dependence of the energy (in arbitrary units $E_{0}$) of the suggested permanently rotating devices as a function 
of the angular frequency $\Omega$ (in units on an optimal frequency $\Omega_{\mathrm{opt}}$). Depending on the design,
the choice of the rotational direction can be either (a) spontaneous or (b) explicit.}
\label{fig:illustrations}
\end{figure}

Suppose for a moment that such an object may indeed be built. What should be the dependence of its energy on the rotational frequency? Qualitatively, one can imagine two physically different situations, depicted schematically in Fig.~\ref{fig:illustrations}. The energy may have two equal minima corresponding to clockwise and counterclockwise directions of rotation, $E(\Omega) = E(-\Omega)$, so that lowest energy state should be chosen spontaneously by the system, Fig.~\ref{fig:illustrations}(a). The energy may also have one global minimum (in addition to possible local minima), so that the reflection symmetry $\Omega \to - \Omega$ may be broken explicitly, Fig.~\ref{fig:illustrations}(b). Below we describe a theoretical proposal of a spontaneously rotating object,  Fig.~\ref{fig:illustrations}(a), although more plausible realizations (in a possible technological sense) may corresponds to the explicit breaking of the macroscopic rotational state, Fig.~\ref{fig:illustrations}(b). 

Philosophically similar ideas were proposed recently in Refs.~\cite{ref:classical:time:crystals,ref:quantum:time:crystals} by A.~Shapere and F.~Wilczek, where a spontaneous breaking of time translation symmetry was suggested in (semi)classical systems~\cite{ref:classical:time:crystals} and in a closed quantum-mechanical system~\cite{ref:quantum:time:crystals}. It was suggested that time symmetry breaking may lead to formation of the ``time crystals'' corresponding to, respectively, (semi)classical motion in the lowest energy state of the system (in a form, for example, of traveling density waves) or to a periodic (in time) reproducibility of the states in a quantum-mechanical system. Here we discuss a possible field-theoretical realization of a permanent rotation due to quantum fluctuations.

In order to illustrate the idea, it is useful first to consider a one dimensional system. The simplest relevant example is given by a vacuum of a $(1+1)$ dimensional theory of a massless real-valued scalar field $\Phi = \Phi(t,\varphi)$ with the Lagrangian $\cL = \frac{1}{2} \partial_{\mu} \Phi \, \partial^{\mu} \Phi$, which is defined on a circle of radius $R$ with angular coordinate $\varphi$.

\begin{figure}[!thb]
\begin{center}
\begin{tabular}{cc}
\includegraphics[scale=0.2,clip=false]{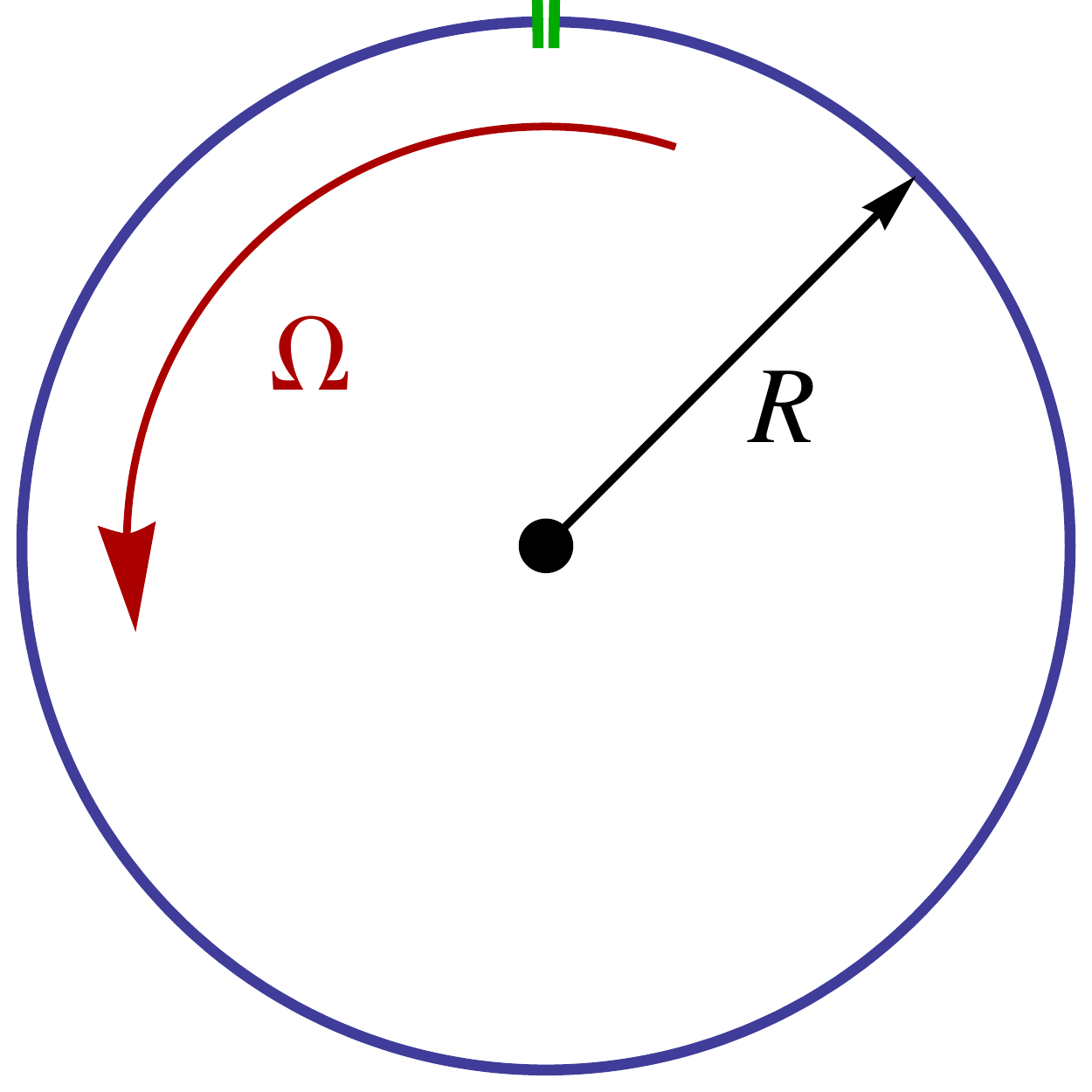} & \hskip 10mm
\includegraphics[scale=0.2,clip=false]{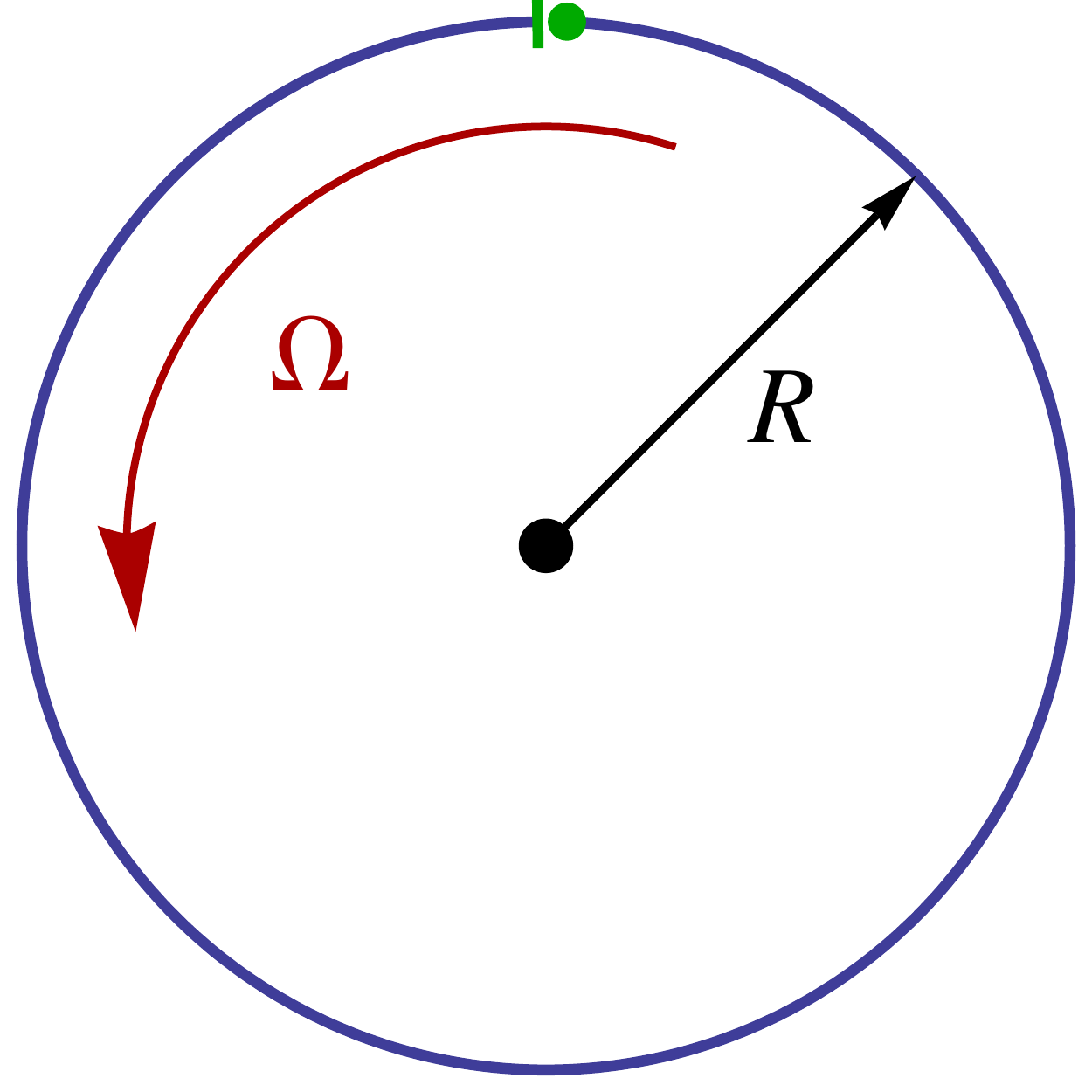} \\[1mm]
(a) & \hskip 10mm (b)
\end{tabular}
\end{center}
\vskip -5mm
\caption{The circles with the cuts which have (a) symmetric (say, Dirichlet-Dirichlet) and (b) asymmetric (say, Dirichlet--Neumann) boundary conditions.}
\label{fig:1d}
\end{figure}

Since the circle is symmetric to rotations about its center we make the system sensitive to rotation by cutting the circle in one point and imposing simplest Dirichlet--type boundary conditions at the both ends of the infinitesimally thin cut, $\Phi(t,\varphi) |_{\varphi \in \mathrm{cut}} = 0$, Fig.~\ref{fig:1d}(a).

Generally, boundary conditions imposed on the quantized fields change the vacuum energy because the quantum fluctuations of the fields (and, thus, their energy spectrum) in the presence of the boundaries are different from the ones in the free space. This is the essence of the celebrated Casimir effect~\cite{ref:Casimir}. In our case, the Dirichlet boundary conditions together with compactness of the circumference of the (non rotating so far) circle gives rise to the energy deficit of the vacuum fluctuations of the scalar field (the Casimir energy), which is known as the L\"uscher energy of an open string of the length $2 \pi R$, with Dirichlet boundary equations at the ends~\cite{Luscher:1980fr}:
\beqn
E^{\mathrm{vac}}_{\mathrm{static}} =\sum_{m=1}^{\infty} \frac{\omega_{m}}{2} \equiv - \frac{1}{48 R}\,.
\label{eq:E:static}
\eeqn
Here $\omega_{m} = m/(2 R)$ (with $m=1,2,\dots$) are the energy levels of the scalar field, and the free-space contribution is subtracted from the divergent sum~\eq{eq:E:static}.

Since the energy of quantum fluctuations~\eq{eq:E:static} is a negative quantity, one can suggest that the moment of inertia $I^{\mathrm{vac}}$, associated with these vacuum fluctuations, should also be negative, implying, in particular, that the vacuum energy of the vacuum fluctuations on the circle should decrease if the circle is set in rotation. Thus, one can suggest that the dependence of the rotational energy on the angular frequency of rotation $\Omega$ should qualitative be similar to the behavior at $\Omega \sim 0$ of the illustrative example of Fig.~\ref{fig:illustrations}(a). One can qualitatively support this suggestion by noticing the fact that the inertial mass corresponding to the Casimir energy $E_{c}$ is always $E_{c}/c^{2}$ regardless of the sign of the Casimir energy~\cite{Fulling:2007xa}, so that the negative mass should have a negative moment of inertia. 

In order to make a quantitative prediction, we calculate the energy of the vacuum fluctuations on our circle, Fig.~\ref{fig:1d}(a), when it rotates uniformly with the angular frequency $\Omega$. The cut of the circumference imposes the following time-dependent boundary condition:
\beqn
\Phi (t,\varphi)\bl_{\varphi = \Omega t}  = 0\,.
\label{eq:circle:D:cond}
\eeqn

The vacuum energy density (averaged over the whole circumference) is given by the following relation~\cite{ref:Milton:Book}:
\beqn
\left\langle T^{00} \right\rangle = \int_{0}^{2\pi} \frac{d \varphi}{2 \pi i}  \partial^{0}\partial^{\prime 0}G(t,t'; \varphi, \varphi)\bl_{t=t'}\,,
\label{eq:E}
\eeqn
where $G(t,t'; \varphi,\varphi')$ is a Green function of the field $\Phi$:
\beqn
\left(\frac{\partial^{2}}{\partial t^{2}} - \frac{1}{R^{2}} \frac{\partial^{2}}{\partial \varphi^{2}} \right) G(t, t', \varphi, \varphi') = \frac{1}{R}\delta(t {-} t') \delta(\varphi {-} \varphi') \,. \quad
\eeqn
The Green function can be expressed as follows:
\beqn
G(t,t';\varphi,\varphi') = \int\limits_{-\infty}^{+\infty} \frac{d \omega}{2 \pi} \sum_{m=1}^{\infty} 
\frac{\Phi^{\dagger}_{\omega, m}(t,\varphi)  \Phi_{\omega, m}(t',\varphi')}{\lambda_{\omega, m} - i \epsilon}, \qquad
\label{eq:G}
\eeqn
where 
\beqn
\Phi_{\omega,m}(t,\varphi) & = & \frac{1}{\sqrt{\pi R}} \sin \Bigl[\frac{m}{2} {[\varphi - \Omega t]}_{2 \pi} \Bigr] \label{eq:phi:m} \\
& & \cdot \exp\biggl\{ - i \omega \biggl(t - \frac{R^{2} \, {[\varphi - \Omega t]}_{2 \pi}}{1 - \Omega^2 R^{2}} \biggr)\biggr\}\,,
\eeqn
is the complex-valued combination of real-valued eigenfunctions (in the laboratory frame) and
\beqn
\lambda_{\omega,m} & = & \frac{m^{2} (1 - \Omega^{2} R^{2})}{4 R^{2}} -\frac{\omega^{2}}{1 - \Omega^{2} R^{2}}\,,
\label{eq:lambda}
\eeqn
are the corresponding eigenvalues (with $\omega \in {\mathbb{R}}$ and $m = 1,2,3, \dots$) of the relevant d'Alembertian:
\beqn
\left(\frac{\partial^{2}}{\partial t^{2}} - \frac{1}{R^{2}} \frac{\partial^{2}}{\partial \varphi^{2}} \right) \Phi_{\omega,m} (t,\varphi) = \lambda_{\omega,m} \Phi_{\omega,m} (t,\varphi)\,,
\eeqn
with the time-dependent Dirichlet boundary condition~\eq{eq:circle:D:cond}. The eigenfunctions~\eq{eq:phi:m} are orthonormalized and their system is complete.

In Eq.~\eq{eq:G}, the choice of the contour in the integration over the frequency $\omega$ (with $\epsilon \to + 0$) corresponds to the Feynman type of Green function~\cite{ref:Milton:Book}.  We would like to notice the presence of the important $2\pi$--modulo operator $[\dots]_{2\pi}$ in Eq.~\eq{eq:phi:m}: $0 < {[x]}_{2 \pi} \equiv x + 2 \pi n < 2 \pi$ with $n \in \Z$. 

Introducing the time--splitting regularization, $t'=t + \delta t$ (which, due to rotation, should be accompanied by the shift of the angular variable $\varphi' = \varphi + \Omega \delta t$ for the sake of consistency), we arrive to the following expression:
\beqn
\left\langle T^{00} \right\rangle & = & \frac{1}{ 2 \pi R (1{-} \Omega^{2} R^{2})} \sum_{m=1}^{\infty}  \int\limits_{-\infty}^{+\infty} \frac{d \omega}{2 \pi i} 
\frac{\omega^{2} {+} {\tilde \omega}_m^2 \Omega^2 R^{2}}{{\tilde \omega}_{m}^{2} {-} \omega^{2} {-} i \epsilon}  e^{- i \omega \delta t} \nonumber 
\nonumber \\
& \equiv &\frac{1}{2 \pi R} \cdot \frac{1 + \Omega^{2} R^{2}}{1 - \Omega^{2} R^{2}} \cdot \sum_{m=1}^{\infty} \frac{{\tilde \omega}_{m}}{2} e^{- {\tilde \omega}_{m} \tau}\,,
\label{eq:T00:1d}
\eeqn
where the effective eigenenergies are:
\beqn
{\tilde \omega}_{m} = m \frac{1 - \Omega^{2} R^{2}}{2 R}\,.
\label{eq:mu:m}
\eeqn
In Eq.~\eq{eq:T00:1d} we have performed a standard Wick rotation~\cite{ref:Milton:Book}, $\delta t \to i \tau$ with $\tau > 0$, and got the regularized expression for the infinite sum over energies.  Next, we notice that the sum in the last line of Eq.~\eq{eq:T00:1d} corresponds to the regularized Casimir energy of the non-rotating circle~\eq{eq:E:static} but with the modified radius ${\tilde R} = R/(1 - \Omega^{2} R^{2})$. Next, one can either set $\tau=0$ and use the $\zeta$ regularization,  $\sum_{m=1}^{\infty} m \equiv \zeta(-1) = - 1/12$, or one can use the known result~\eq{eq:E:static} with the replacement $R \to {\tilde R}$, and get the following result for the energy $E^{\mathrm{vac}} \equiv 2 \pi R \langle T^{00} \rangle$ of the rotating circle with the Dirichlet cut, Fig.~\ref{fig:1d}(a):
\beqn
E^{\mathrm{vac}} (\Omega) = - \frac{1 + \Omega^{2} R^{2}}{48 R}\,.
\label{eq:E:vac:1d}
\eeqn
At $\Omega = 0$ the vacuum energy reduces to the static expression~\eq{eq:E:static}. Equation~\eq{eq:E:vac:1d} can (naively) be interpreted with an intuitive help of the classical formula~\eq{eq:E:class} in terms of the (negative) vacuum moment of inertia, $I^{\mathrm{vac}} = - R/24$. 

A real physical device of the discussed geometry should have a large classical moment of inertia, and the classical kinetic energy of rotation~\eq{eq:E:class} should definitely overcome the quantum contribution~\eq{eq:E:vac:1d}. Notice that in the physical world, the rotational instability~\eq{eq:E:vac:1d} may only be valid at small frequencies due to obvious deformations of physical structures at large $\Omega$. 

Can we propose a real physical object, for which a rotational energy is minimized at a nonzero rotational frequency? It may only be possible is the quantum fluctuations violate the reflection symmetry $\Omega \to - \Omega$ explicitly, as suggested in Fig.~\ref{fig:illustrations}(b). Then the sum of the symmetric classical rotational energy~\eq{eq:E:class} of the object and its asymmetric quantum energy, Fig.~\ref{fig:illustrations}(b), should have a minimum at nonzero angular frequency regardless of the mass of the object. One may suggest that the desired effect may be reached if the boundary conditions at different ends of the cut are inequivalent. For example, one can imagine the Dirichlet condition at one end and a Neumann condition the other end, as shown in Fig.~\ref{fig:1d}(b). Then the quantum fluctuations should be sensitive to the direction of the rotation, and the vacuum energy should not be symmetric under reflections $\Omega \to - \Omega$. Such structure will prefer to rotate forever.

\begin{figure}[!thb]
\begin{center}
\includegraphics[scale=0.4,clip=false]{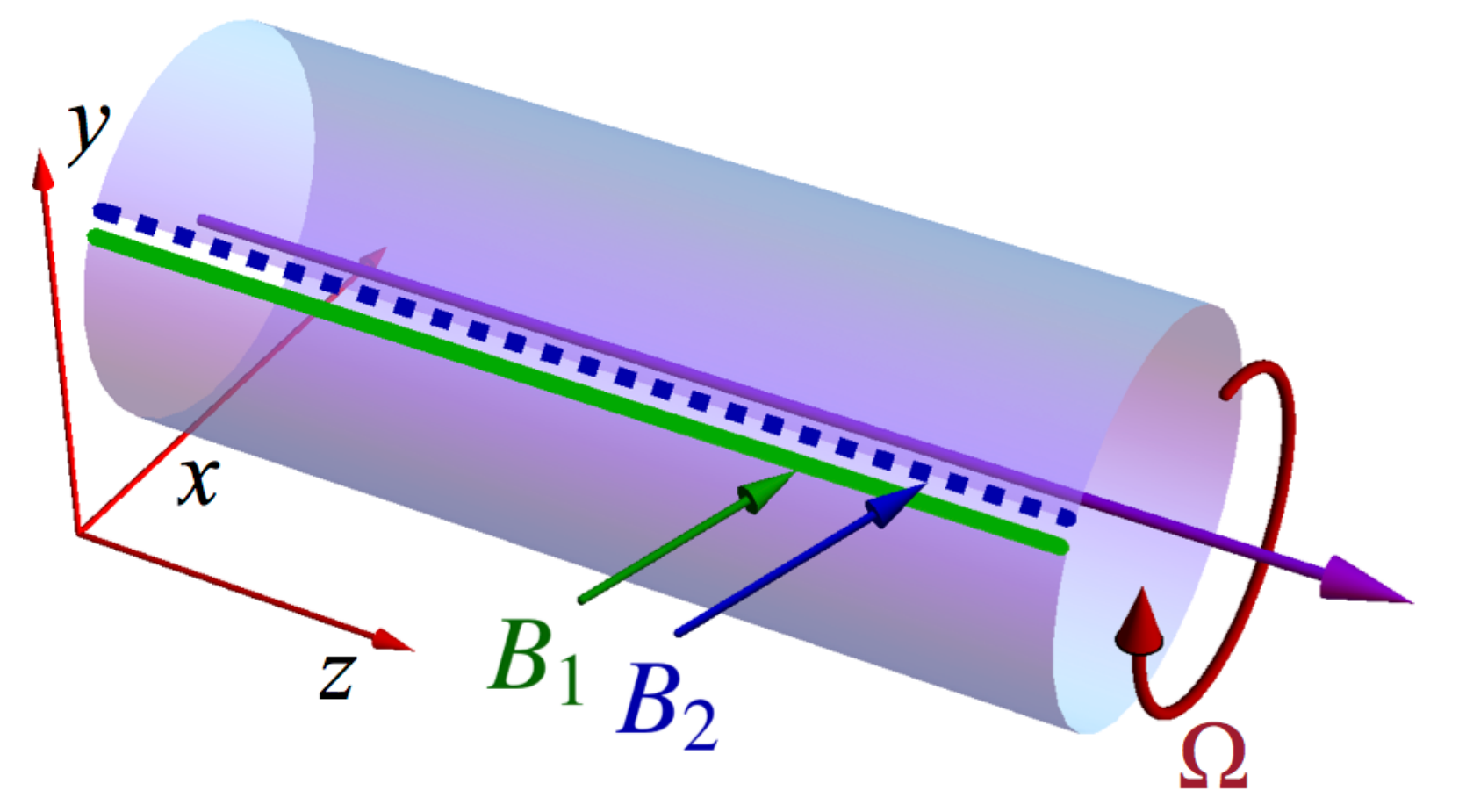}
\end{center}
\vskip -5mm
\caption{The two dimensional setup (described in the text).}
\label{fig:2d}
\end{figure}

A two-dimensional analogue of the discussed systems is given by a tube of radius $R$ and height $L$, which has infinitesimal cut along the $z$ axis, Fig.~\ref{fig:2d}. Here we assume that the boundary conditions $B_{1,2}$ at the sides of the cut and at other edges of the tube are of the Dirichlet type.

The eigenfunctions of the tube with the cut are 
\beqn
\Phi^{(2d)}_{\omega,m,n}(t,\varphi,z) = \Phi_{\omega,m}(t,\varphi,z) \cdot \sqrt{\frac{2}{L}}\sin \frac{\pi n z}{L}\,,
\eeqn
where the longitudinal quantum number is $n = 1,2,\dots$ and the polar wavefunction $\Phi_{\omega,m}$ is given in Eq.~\eq{eq:phi:m}.
The corresponding eigenvalues are $\lambda^{(2d)}_{\omega,m,n} = \lambda_{\omega,m} + \left(\frac{\pi n}{L} \right)^{2}$, where $\lambda_{\omega,m}$ is given in Eq.~\eq{eq:lambda}.

Following our previous line of considerations, we get the vacuum energy of the rotating tube, Fig.~\ref{fig:2d}:
\beqn
E^{\mathrm{vac}}_{\mathrm{tube}} (\Omega) = \frac{l}{\pi R} \left(1 - \Omega^{2} R^{2} l \frac{\partial}{\partial l}\right) E^{\mathrm{vac}}_{\mathrm{rect}} (L,l) \bl_{l = l_{0}(R,\Omega)}\,, \qquad
\label{eq:E:vac:tube}
\eeqn
where $l_{0}(R,\Omega) = 2 \pi R/ \sqrt{1 - \Omega^{2} R^{2}}$. Here
\beqn
E^{\mathrm{vac}}_{\mathrm{rect}} (L,l) & = & \frac{\pi}{2} \sum_{m,n=1}^{\infty} {\left[ {\left(\frac{n}{L} \right)}^{2} +  {\left(\frac{m}{l} \right)}^{2} \right]}^{\frac{1}{2}}\nonumber \\
& \overset{\text{ren}}{=} &\frac{\pi}{48 L} - \frac{\zeta(3) l}{16 \pi L^{2}} + \frac{\pi}{L} G\left( \frac{l}{L} \right)\,,
\label{eq:E:vac:Ll}
\eeqn
is the Casimir energy of a massless scalar field in $L \times l$ rectangle with the Dirichlet boundaries~\cite{Ambjorn:1981xw,ref:Bordag:Book}. The second line in Eq.~\eq{eq:E:vac:Ll} is the renormalized energy with
\beqn
G(z) = - \frac{1}{2 \pi} \sum_{n,m = 1}^{\infty} \frac{n}{m} K_{1} (2 \pi n m z)\,,
\eeqn
and $K_{1}(z)$ is the modified Bessel function. The details of the calculation will be presented elsewhere.

\begin{figure}[!thb]
\begin{center}
\includegraphics[scale=0.34,clip=false]{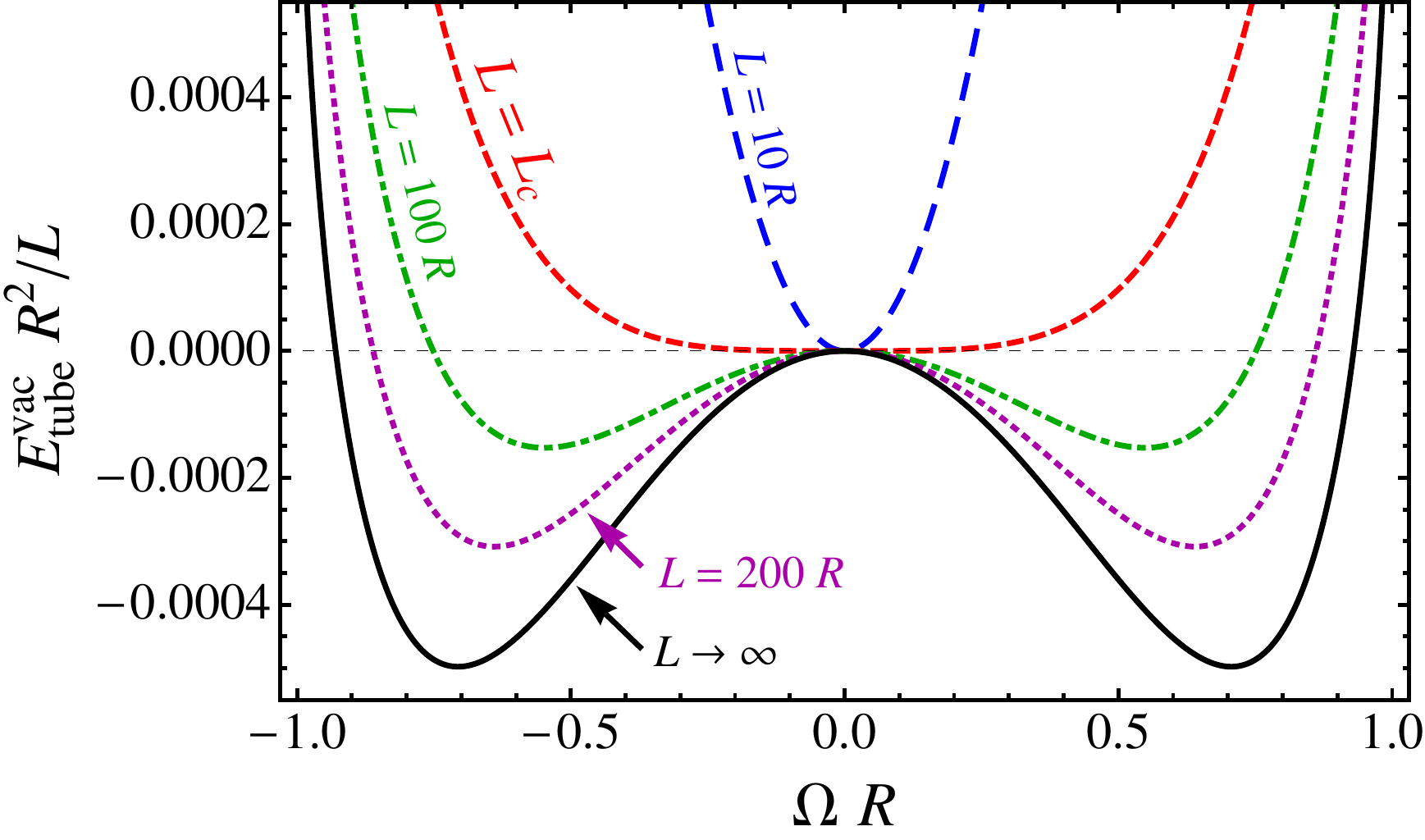}
\end{center}
\vskip -5mm
\caption{Vacuum energy~\eq{eq:E:vac:tube} of the rotating tube, Fig.~\ref{fig:2d},~of radius $R$ and height $L$, as a function of the angular frequency~$\Omega$.}
\label{fig:E:Omega}
\end{figure}

The rotational contribution to the vacuum energy~\eq{eq:E:vac:tube} is shown in Fig.~\ref{fig:E:Omega} (with subtraction of an $\Omega$--independent term). It takes a simpler form for a long tube,
\beqn
E^{\mathrm{vac}}_{\mathrm{tube}}  \bl_{\frac{L}{R} \gg 1} =  \frac{\zeta(3)L}{32 \pi^{3} R^{2}} \left[1 - \sqrt{1 - R^{2} \Omega^{2}} \left(1 + 2 R^{2} \Omega^{2} \right) \right]\,.\nonumber
\label{eq:Elong:quant}
\eeqn

\begin{figure}[!thb]
\begin{center}
\includegraphics[scale=0.34,clip=false]{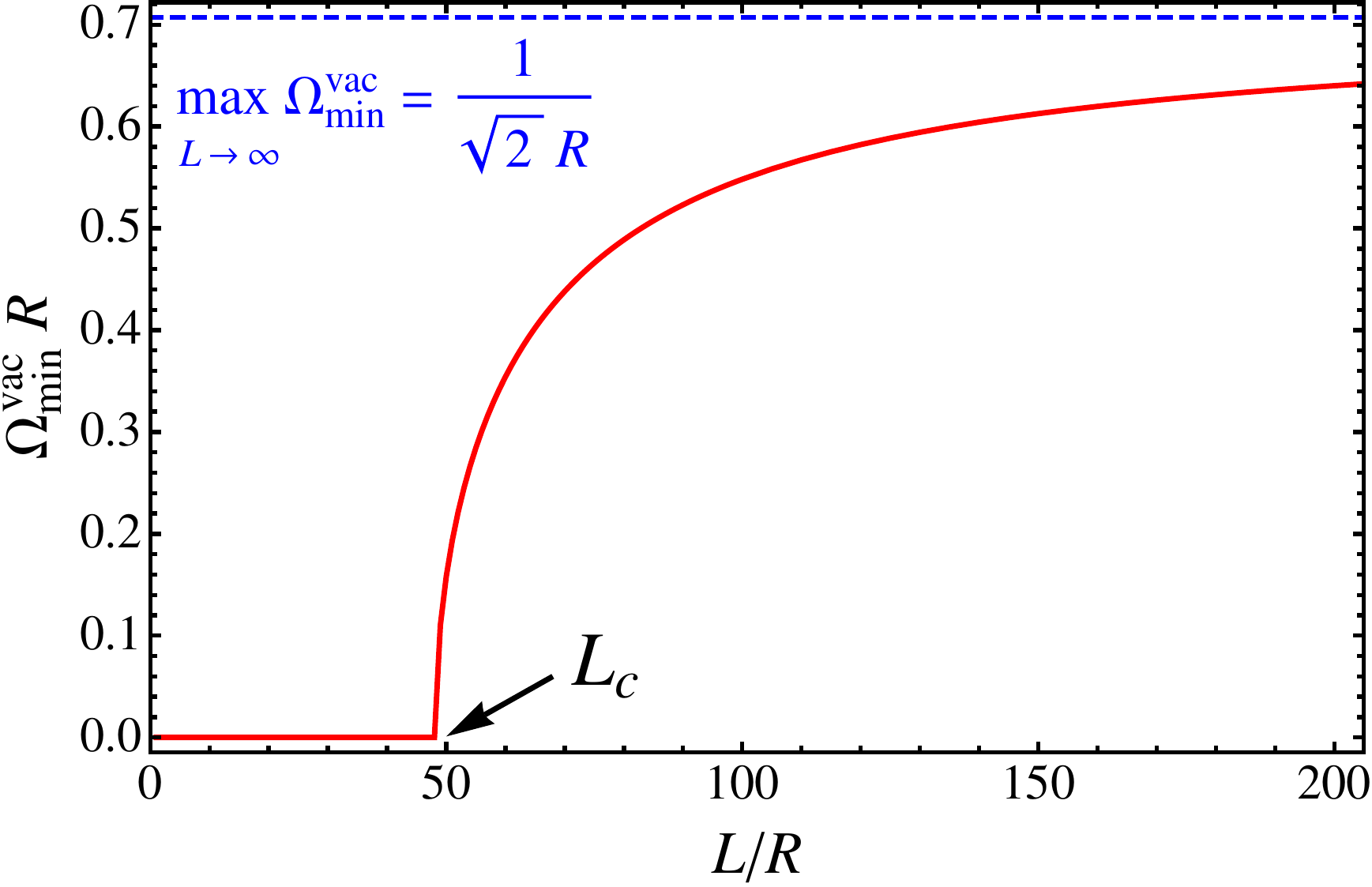}
\end{center}
\vskip -5mm
\caption{The angular frequency at which the minimum of the vacuum energy~\eq{eq:E:vac:tube} is reached vs the height of the tube $L$.}
\label{fig:Omega:min}
\end{figure}

If the height of the tube exceeds the critical length
\beqn
L_{c} (R) \approx 48.1 \, R\,,
\label{eq:Lc}
\eeqn
the energetically favorable state of vacuum fluctuations corresponds to a rotation of the tube with a nontrivial minimum at $\Omega = \pm \Omega^{\mathrm{vac}}_{\mathrm{min}} (L) \neq 0$, Fig.~\ref{fig:Omega:min}.

In the long--tube limit the energy of quantum fluctuations~\eq{eq:Elong:quant} has a nontrivial minimum at
\beqn
\Omega = \lim_{L\to\infty}\Omega^{\mathrm{vac}}_{\mathrm{min}} (L) \equiv \frac{1}{\sqrt{2}} \frac{1}{R}\,.
\label{eq:Omega:min:long}
\eeqn
If $L < L_{c}$, then the energy minimum is trivial, $\Omega^{\mathrm{vac}}_{\mathrm{min}} = 0$.

At $\Omega = 0$ the curvature of the quantum rotational energy~\eq{eq:Elong:quant} is a quadratic function of the angular frequency $\Omega$ similarly to the classical expression~\eq{eq:E:class}. The corresponding moment of inertia per unit height of the long tube is a negative quantity,
\beqn
\lim_{L \to \infty} \frac{I^{\mathrm{vac}}(\Omega=0)}{L} = - \frac{3 \zeta(3) \hbar}{32\pi^{3} c} \approx -0.003635 \, \frac{\hbar}{c}\,,
\label{eq:rho:fl:z}
\eeqn
which is a universal number independent of the radius of the tube. It depends only on the type of the boundary conditions imposed along the cut.

The behavior of the vacuum energy, Fig.~\ref{fig:E:Omega}, corresponds to the spontaneous rotation, as it was illustrated in Fig.~\ref{fig:illustrations}(a). In order to avoid the suppression of the quantum rotational effect by the large classical contribution~\eq{eq:E:class}, one can make the rotational symmetry breaking explicit, Fig.~\ref{fig:illustrations}(b), by imposing different boundary conditions $B_{1}$ and $B_{2}$ at the edges of the cut in Fig~\ref{fig:2d}. 

The quantum rotational energy of a realistic massive tube with asymmetric boundary conditions should be extremely small compared to its classical counterpart~\eq{eq:E:class}. At the same time, the massless tube, Fig.~\ref{fig:2d}, has an unbelievably huge, relativistic optimal vacuum frequency~\eq{eq:Omega:min:long}, $\Omega \sim c/R$, where $c$ is the speed of light. In real (massive) asymmetric devices the optimal frequency should be determined by a competition of these two opposite factors.

One can imagine a realization of the proposed designs with (carbonic/metallic) nanotubes or with larger objects made of, for example, graphene which is known to host massless fermionic excitations~\cite{Novoselov:2005kj} (indeed, the Casimir effect is also present in compact systems with either massless or massive fermions~\cite{ref:fermionic}).

\begin{figure}[!thb]
\vskip -3mm
\begin{center}
\includegraphics[scale=0.5,clip=false]{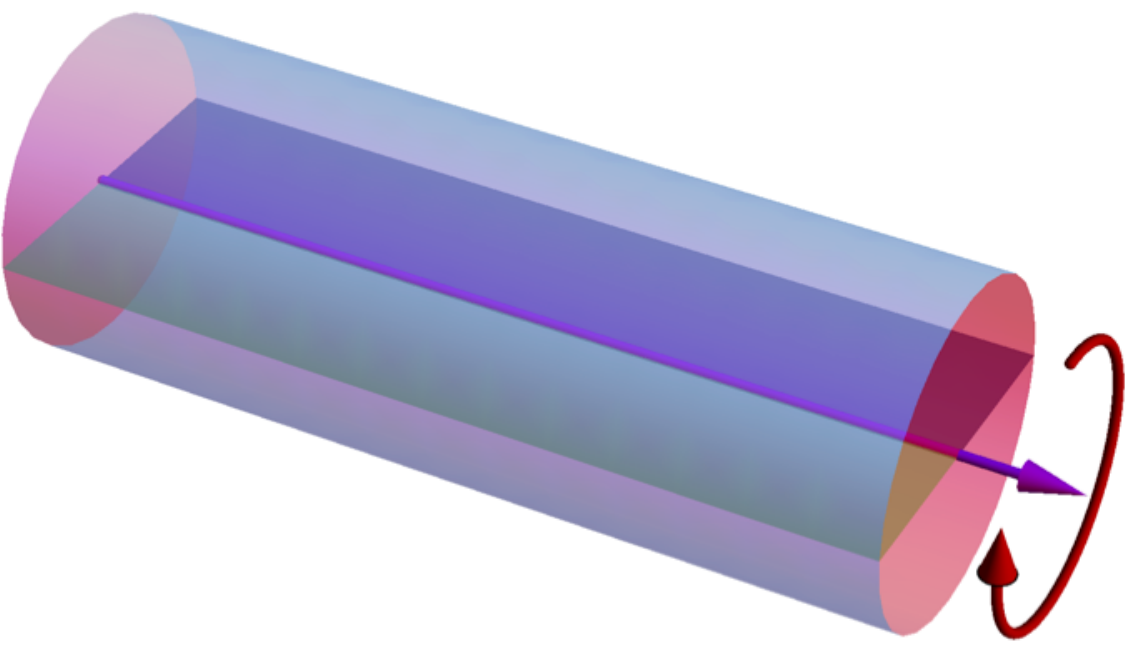}
\end{center}
\vskip -5mm
\caption{A generalization of the rotating device to three dimensions (the description is provided in the text).}
\label{fig:3d}
\end{figure}

As a possible generalization of the mentioned one- and two-di\-men\-sio\-nal devices to three dimensions, one can suggest a cylinder with a central bar, Fig.~\ref{fig:3d}. The central bar should have two different, preferably, double-sided parts, made of electrically conducting materials with, generally, different properties at each side. Such device should be sensitive to the direction of rotation. The ``Rotational Vacuum Effect'' should be mediated by the vacuum quantum fluctuations of the electromagnetic field, as it happens in the standard Casimir effect~\cite{ref:Casimir}. 

Is the rotation in the electromagnetic vacuum dissipative, can it last forever? In fact, a single spinning object should experience a rotational friction (the ``vacuum friction''~\cite{ref:Manjavacas}) which may eventually slow the object down due to radiation. It was stressed in Ref.~\cite{Maghrebi:2012tv} that the rotating object should spontaneously emit radiation only for the so-called superradiating (Zel'dovich) modes~\cite{ref:Zeldovich}. The quantization of the emitted radiation may forbid dissipation (as it happens, say, in atoms where the quantization of electrons eigenstates prohibits continuous radiation by the orbiting electrical charges thus providing stability to lowest electron states in the atoms).

For completeness, we would like to mention a possible weak point of our calculations which is related to the renormalization of the quantum fluctuations. Indeed, in our derivation we used the known results obtained in the static systems (the Casimir energies corresponding to an open string and to a rectangle), while the renormalization in a non-inertial frame of rotating systems may provide quantitative corrections to our results.

The very existence of such permanently rotating device is very difficult to accept intuitively. The suggested device is, however, not a perpetuum mobile: the system rotating with optimal frequency is already in its lowest energy state and it is impossible to extract energy from this rotation. Moreover, in order to slow down (or, to stop completely) the rotation, one needs to input certain energy to the system. This excess in energy may subsequently be released back (possibly, in a form of radiation or as a mechanical recoil, depending on a physical setup).

Finally, we notice that the proposed structures with asymmetric boundary conditions should have different moments of inertia in the clockwise and counterclockwise directions of rotation. Thus, the Rotational Vacuum Effect has a potential to be observed experimentally. 

The work was supported by Grant No. ANR-10-JCJC-0408 HYPERMAG.

\end{document}